\documentclass[lettersize,journal]{IEEEtran}
\usepackage{amsmath,amsfonts}
\usepackage{algorithmic}
\usepackage{algorithm}
\usepackage{array}
\usepackage[caption=false,font=normalsize,labelfont=sf,textfont=sf]{subfig}
\usepackage{textcomp}
\usepackage{stfloats}
\usepackage{url}
\usepackage{verbatim}
\usepackage{graphicx}
\usepackage[numbers]{natbib}

\usepackage{longtable}
\usepackage{array}

\usepackage{booktabs}
\usepackage{multirow}
\usepackage[switch]{lineno}
\usepackage{amsfonts,bbm,bm,courier}
\usepackage{placeins}

\newcommand{\x}{\bm{x}}
\newcommand{\cf}{\bm{c}}

\newcommand{\citeN}[1]{\citeauthor{#1} \cite{#1}}

\begin{document}

\title{An Actionability Assessment Tool for Explainable AI}

\author{Ronal Singh, Tim Miller, Liz Sonenberg, Eduardo Velloso, Frank Vetere, Piers Howe, Paul Dourish
\thanks{Dr Ronal Singh is with CSIRO's Data61, Melbourne, Australia}
\thanks{Prof Tim Miller is with the University of Queensland, Brisbane, Australia}
\thanks{Prof Liz Sonenberg, Prof Frank Vetere, and A/Prof Piers Howe are with the University of Melbourne, Melbourne, Australia}
\thanks{Prof Eduardo Velloso is with the University of Sydney, Sydney, Australia}
\thanks{Prof Paul Dourish is with the University of California, Irvine, US}
}

\maketitle

\pagenumbering{gobble}

\begin{abstract}
In this paper, we introduce and evaluate a tool for researchers and practitioners to assess the actionability of information provided to users to support algorithmic recourse. While there are clear benefits of recourse from the user's perspective, the notion of actionability in explainable AI research remains vague, and claims of `actionable' explainability techniques are based on the researchers' intuition. Inspired by definitions and instruments for assessing actionability in other domains, we construct a seven-question tool and evaluate its effectiveness through two user studies. We show that the tool discriminates actionability across explanation types and that the distinctions align with human judgements. We show the impact of context on actionability assessments, suggesting that domain-specific tool adaptations may foster more human-centred algorithmic systems. This is a significant step forward for research and practices into actionable explainability and algorithmic recourse, providing the first clear human-centred definition and tool for assessing actionability in explainable AI.
 
\end{abstract}


\section{Introduction}
\emph{Algorithmic recourse} is the provision of an explanation of an algorithmic decision to support an individual to determine \emph{actionable} steps they can take to alter a decision -- typically focussed on decisions the user perceives as unfavourable~\cite{karimi2022,Sullivan2022-sm}. 
When AI systems make decisions that significantly affect people's lives, the agents involved in human-agent interaction must be capable of generating such actionable explanations.

While there are clear benefits of recourse from the user's perspective, there is no agreement on the definition of `actionability' in the context of explanations, nor of how to assess it \cite{karimi2022,rasouli2022,Kelechi2023-aw}. 
Many research papers claim that certain explainable AI approaches are `actionable' using a technical definition of actionability, e.g. \cite{karimi2022,Ustun2019-ic}, and actionability is then typically assessed using only the researchers' intuition as to what constitutes an actionable response. As warned by \citeN{Leavitt2020-kj} and \citeN{Miller2017-cu}, intuition-based assessment leads to ``illusory progress and misleading conclusions'' \cite{Leavitt2020-kj}. 

In this paper, we aim to bridge a critical gap in XAI research by proposing and validating an \textit{Actionability Assessment Tool for XAI} for researchers and practitioners to assess the actionability of an explanation provided for algorithmic recourse. This provides a human-centred criterion allowing the actionability of explanations to be tested without relying solely on researcher or practitioner intuitions. 

We reviewed existing instruments related to actionability from different fields and identified five topics and seven questions as candidates for the final version of the tool. Inspired by the definition used in the domain of patient education~\cite{Shoemaker2014-wo}\footnote{\url{https://www.ahrq.gov/health-literacy/patient-education/pemat1.html}}, we define `actionability' as follows:

\begin{quote}
	\emph{Definition}: An explanation of a decision is \emph{actionable} if people can use the information to identify actions to take to change the decision.
\end{quote}

To contribute to the growing understanding of how AI-generated explanations can be useful ~\cite{Byrne-ijcai2023survey,Hoffman2023a}, we conducted two studies to evaluate the tool's capacity to distinguish the actionability of various explanation types. We ran experiments in two domains, credit scoring (deciding whether to approve or deny loans) and employee turnover (deciding whether an employee was likely to resign), and with three example-based explanations: prototypical ~\cite{Molnar2020-fj,Gurumoorthy2019-oe}, counterfactual~\cite{Wachter2017-fv,Russell2019-jh} and directive~\cite{Singh2023-is}. The purpose of the studies was not to rate the actionability of specific types of algorithmically generated explanations but to evaluate the actionability tool itself. Since our subjects were presented with hypothetical scenarios and were not in decision-making settings with the opportunity to take real steps towards recourse, the ratings correspond to \emph{perceived} actionability. 

The explanation types we used were all example-based and were chosen as we anticipated they would display meaningfully different levels of actionability. 
Specifically, we hypothesised that human subjects would rate prototypical explanations as the least actionable, followed by counterfactual explanations, with directive explanations rated the most actionable.

We conducted two studies: one in which participants rated the actionability of explanations directly based on individual judgement, providing evidence as to which of the tested explanations were perceived to be most actionable (Study 1), and one in which a different set of participants used the actionability assessment tool to rate the same explanations (Study 2). Evidence for the tool's value was derived from the demonstrated alignment between these two sets of ratings.  

\IEEEpubidadjcol

Our findings highlight the tool's ability to differentiate among explanation types and that this differentiation aligns with the judgement of human subjects. The study reveals that participants found directive explanations to be the most actionable, as they clearly outlined specific steps. The research also underscores the influence of contextual factors on explanation ratings, revealing how participants' roles significantly shape their perceptions. This indicates that adapting the tool to specific application domains may be useful. This is an important step to better falsifiable research in explainable AI. This work contributes valuable insights towards designing more human-centred algorithmic systems by demonstrating the tool's effectiveness in distinguishing actionability and highlighting the importance of context.

\section{Literature Review}
This section highlights the challenges of creating actionable explanations. The review also examines actionability tools from various domains, highlighting their potential in assessing explanation actionability in XAI.

\subsection{Counterfactual, prototypical and directive explanations}
\emph{Counterfactual explanations} describe ``how the world would have to be different for a desirable outcome to occur'' and have the potential to enable recourse~\cite{Wachter2017-fv}. 
They are typically generated by identifying minimal changes in the input features that lead to the desired outcome from the model.
Various criteria for `minimality' are used; for example, some researchers highlight proximity~\cite{Wachter2017-fv},  some suggest sparsity~\cite{Mothilal2020-qi}, and others diversity~\cite{Russell2019-jh}. Recent reviews include \citeN{karimi2022,Aryal-Keane-ijcai2023survey}.

By raising awareness of similar features that result in the desired prediction, counterfactual explanations may guide what circumstances would result in a different outcome. However, they do not necessarily indicate feasible actions that may lead to this desired result; that is, they do not provide explicit recommendations relevant to the user on \textit{how to act}~\cite{karimi2022,Singh2023-is}, so they offer limited agency for the individual seeking recourse. Similar sentiments have been raised recently by researchers within the AAMAS community~\cite{Leofante2023-dt,Gajcin2023-eg}. 

Researchers propose offering multiple diverse counterfactual explanations with the hope that at least one is suitable for the recipient \cite{Ustun2019-ic,Mothilal2020-qi,Poyiadzi2020-tp}. Others propose optimisation methods where domain knowledge and user preferences can be incorporated into generating recommendations \cite{rasouli2022}, and some limit the variables to domain-specific actionable features \cite{bhattacharya2023directive}. 

\textit{Prototypical explanations} present instances from the dataset to explain the model's behaviour. These are prototypes:  instances representative of all the data~\cite{Molnar2020-fj}. Our study uses them because they are example-based, similar to counterfactuals.

\textit{Directive explanations} have been initially proposed by \citeN{Singh2023-is}: they are explanations that provide recommendations of the action(s) the individual could perform to achieve the counterfactual state. Directive explanations have been studied in an abstract setting with na\"{i}ve users \cite{Singh2023-is} and in a health setting with expert users \cite{bhattacharya2023directive}. Related considerations have been proposed for generating intervention recommendations when causal models are available \cite{karimi2022,Karimi2021-ke}.

\subsection{Actionability instruments}\label{sec:instruments}

Actionability has been considered in several domains including data mining~\cite{Wang2002-ys}, emotion mining~\cite{Tzacheva2020-kt}, cybersecurity~\cite{Redmiles2020-jb} and the development of educational materials~\cite{Shoemaker2014-wo}. Some have existing tools or instruments used by industry or researchers. This section describes several existing actionability measurement tools used from domains outside AI.

\subsubsection{The patient education materials assessment tool}
The Patient Education Materials Assessment Tool (PEMAT)\footnote{https://www.ahrq.gov/health-literacy/patient-education/pemat.html} is a systematic method to evaluate and compare the understandability and actionability of patient education materials~\cite{Shoemaker2014-wo}. It was designed as a guide to help determine whether patients can understand and make informed decisions about what they can do to improve their health outcomes based on the provided learning materials and information. The intent of PEMAT significantly overlaps with the aim of algorithmic recourse, making PEMAT relevant for discussions on what an actionability tool would look like for the XAI. 

\subsubsection{Actionability of management research}
HakemZadeh and Baba \cite{HakemZadeh2016-ba} developed an instrument to assess the actionability of management research using the following criteria.
\textit{Operationality}: legal and cost-effective.
\textit{Causality:} establish a cause-and-effect relationship; that is, a causal pathway, between a decision and its decision outcomes.
\textit{Contextuality:}
incorporate the effects of contextual factors on the decision and decision outcome~\cite{Syed2010-bf}.
\textit{Comprehensiveness:} be sufficiently comprehensive and capture the complexity and dynamics of the decision processes and diversity of knowledge.
\textit{Persuasiveness:} demonstrate face validity and persuade the decision maker to implement the findings. 
\textit{Comprehensibility:} understood easily.
\textit{Conceptual clarity:} offer a logical mental model to managers to help them with decision options~\cite{Converse1993-uo}. These factors also resonate in the discussions with the algorithmic resource community. Therefore, the above instrument is relevant to judging the actionability of XAI. 

\subsubsection{Relevant instruments from shared decision-making literature}
\citeN{Scholl2011-gp} review several instruments for measuring various aspects of shared decision-making (SDM) involving multi-party decisions. These instruments measure the quality of information or evidence used to make such decisions. We found two relevant instruments: the \textit{Satisfaction with Decision Instrument}, which measures patients' satisfaction with medical decisions~\cite{Sainfort2000-ta}, and the \textit{Decision self-efficacy scale} used by \cite{OConnor1995-yb} to measure a patient's self-confidence in making decisions.

\subsubsection{Perceived Actionability of Security Advice Available on the Web}

Redmiles et al. \cite{Redmiles2020-jb} assess the perceived actionability of security advice using four sub-metrics: \emph{Confidence}, \emph{Time Consumption}, \emph{Disruption}, and \emph{Difficulty}. These sub-metrics were developed considering theories relevant to security behaviour, such as the Protection Motivation Theory \cite{Rogers1997-pv}, the Human in the Loop model \cite{Cranor2008-bl}, and the economic frameworks of secure behaviour~\cite{Redmiles2018-kq}. 

\section{Actionability Assessment Tool}
To develop a novel tool for measuring actionability specific for the context of XAI/recourse,
we first reviewed the five actionability instruments described in Section~{\ref{sec:instruments}: ~\cite{Shoemaker2014-wo,HakemZadeh2016-ba,OConnor1995-yb,Sainfort2000-ta,Redmiles2020-jb}. From these, we synthesised questions and topics that might enable the assessment of actionability for an explanation that could enable algorithmic recourse.

Initially, we identified eight topics covered by a total of 22 questions. We then judged the relevance of the topics and questions to the requirements for XAI. Questions that overlapped in terms of what they were assessing were dropped or amended, and we phrased the questions from the user's perspective. For example, rather than asking whether an explanation provides an action, we ask whether \textit{the user} could identify an action using the explanation. After several iterations, we identified five topics and seven questions to be used in the version of the tool to be studied empirically in the XAI context.

Next, we present the topics and questions we included in our  \textit{Actionability tool for XAI}. Although the tool aims to evaluate the actionability of \textit{explanations}, to increase its generality and to avoid biasing participants, all questions were phrased to refer to the ``information'' the participant received.

\subsection{Clarity of information}
Management research literature deems comprehensibility of the findings important, that is, \textit{research should be understood easily}~\cite{HakemZadeh2016-ba}. When designing PEMAT, the researchers considered how different types of information make it easier for patients to act on the information, such as visual aids, tools and instructions on how to use these~\cite{Shoemaker2014-wo}. Our question judges whether the actionable information is easy to understand. The information should use easy-to-understand language to make it easy for people to use. The following item assesses the presentation of the information:

\begin{quote}

\begin{itemize}
    \item[\textbf{Q1}] The information is clear and easy to understand. 
\end{itemize}

\end{quote}

\subsection{Decision understanding}
Existing instruments assess whether people can identify the reasons for the decision. For example, instruments assessing the actionability of management research highlight that actionable information should \textit{explain the reasons}~\cite{HakemZadeh2016-ba} for the decision. Similarly, PEMAT assesses actionability and understandability using different sets of items, but PEMAT assesses \textit{both}~\cite{Shoemaker2014-wo}. Finally, the shared decision-making instruments emphasise helping patients get facts about the decisions, risks and benefits of choices~\cite{OConnor1995-yb}. \cite{Sullivan2022-sm} \cite{Sullivan2022-sm} argue that it is essential that actionable information provides an understanding of the decision. If it does not, the information only provides a recommendation with no actionable information. To assess whether the explanation provides people with an understanding of the reasons for the decision, we include the following question:

\begin{quote}
\begin{itemize}
    \item[\textbf{Q2}] The information helps me understand the reason(s) for the decision.
\end{itemize}
\end{quote}

\subsection{Personalisation/relevance/contextualisation}
Actionable information is subjective and depends on the stakeholder, the domain, and the user's specific goals. The subjectiveness is captured in different ways in existing instruments. For example, PEMAT assesses whether the information provided to patients \textit{directly addresses} them. The instruments in shared decision-making suggest that actionable information must help people identify actions that suit them and align with their values~\cite{Sainfort2000-ta,OConnor1995-yb}. Management research literature indicates that research is made actionable by incorporating and explaining how the situational factors affect the decision, decision outcome and course of action~\cite{Syed2010-bf,HakemZadeh2016-ba}. 

We want actionable information to be relevant to the recipient and the specific circumstances. Information should be socially appropriate, i.e., align with people's values~\cite{Sainfort2000-ta} and should not be \textit{condescending}~\cite{Singh2023-is}. The following two questions assess the relevance of the information to the user and whether the information is socially appropriate:

\begin{quote}
\begin{itemize}
    \item[\textbf{Q3}] The information is relevant to my personal circumstances.
    \item[\textbf{Q4}] The information is socially appropriate.
\end{itemize}
\end{quote}

\subsection{Correction of misunderstandings}
People should be allowed to present arguments, clarifications or other information. For example, when they recognise that the algorithm may have used incorrect information~\cite{Venkatasubramanian2020-vz}. Instruments in shared decision-making, for example, have multiple items that assess whether patients can ask questions, express their views, or ask for advice~\cite{OConnor1995-yb}. We have the following question to capture whether the information helps people correct misunderstandings:
   
\begin{quote}
\begin{itemize}
    \item[\textbf{Q5}] The information allows me to identify and correct any misunderstandings of my personal situation.
\end{itemize}
\end{quote}

\subsection{Action}
Across all actionability instruments, we find explicit references to \textit{actions}. For example, PEMAT assesses whether information \textit{clearly identifies at least one action} for the patients~\cite{Shoemaker2014-wo}. Management research instruments judge whether findings \textit{recommend a cost-effective action}~\cite{HakemZadeh2016-ba}, and the cybersecurity instrument judges how difficult and time-consuming it would be to implement the recommendations~\cite{Redmiles2020-jb}. Similarly, PEMAT assesses whether information helps patients break down information into steps. Actionable information should allow people to identify at least one feasible action to overturn a (negative) decision and to identify steps to execute. The following questions assess these aspects.

\begin{quote}
\begin{itemize}
    \item[\textbf{Q6}] The information allows me to identify at least one feasible action to achieve my desired outcome.
    \item[\textbf{Q7}] The information allows me to break down any action into explicit steps. 
\end{itemize}
\end{quote}

\section{Study Design}
Our two studies evaluated the effectiveness of the proposed actionability tool over decision scenarios in two domains (credit domain and employee domain) and on three explanation types: prototypical explanations~\cite{Gurumoorthy2019-oe}, counterfactual explanations~\cite{Wachter2017-fv}, and directive explanations~\cite{Singh2023-is}. We hypothesised that prototypical explanations would be rated the least actionable, followed by counterfactual explanations, with directive explanations the most highly rated.

It is important to note that the study aims to evaluate the actionability tool, \emph{not} any particular explainability methods. We chose directive, counterfactual and prototypical explanations as three types of explanation that we anticipated would display meaningful differences in actionability. As these three methods all involve example-based explanations, we could control for factors such as explainability presentation. Substituting one of these for a largely different type of explainability method, such as feature-based explanations, would make it difficult to identify whether differences in the actionability ratings were due to the presentation of the explanation or other factors, such as the explanation's understandability.

We conducted two simultaneous within-subjects studies with two different participant groups. 
In Study 1, one set of participants received three scenarios and two explanations of different types per scenario. The participants were asked to rate which explanation in the pair was more `actionable'. In Study 2, a different set of participants received three scenarios, one explanation type per scenario. They rated the explanation using the tool. As evidence for the tool's ability to discriminate actionability, we looked for alignment between the relative order of actionability of explanations reported by participants (Study 1) and the ratings allocated by participants using the tool (Study 2). The explanation types are the independent variables. The dependent variables are the pairwise selections (Study 1) and the question ratings (Study 2). 

\subsection{Scenarios}
We designed four lending decision scenarios in which a machine learning (ML) model approved or denied loan applications. Similarly, we designed four scenarios around employee turnover decisions in which an ML model predicted whether an employee would resign or stay in a company. For each domain, two scenarios were favourable (e.g., the loan was approved), and two were unfavourable (the loan was denied). We used the same scenarios for the two studies.  

We presented participants with an employee/loan applicant profile in a tabular format showing features used by the machine learning model to make predictions. The participants also received a textual description detailing the features contributing to the decision. Table~\ref{tab:sample_employee_profile} shows an example of an employee profile from the employee domain\footnote{Please see the appendix for a complete list of scenarios}.

\begin{table}[!ht]
    \small
    \centering
    \caption{Sample Employee Profile}
    \begin{tabular}{p{5cm}|p{2.4cm}}
        \toprule
        \multicolumn{2}{l}{Employee: Tanya} \\
        \midrule
        \textbf{Feature} & \textbf{Value} \\
        \midrule
        Age & 43 \\
        Business Travel & Medium \\
        Employment length (in the company)  & 4 years \\
        Employment length (lifetime)	& 7 years \\
        Months since last promotion	& 12 \\
        Monthly income (\$)	& 5300 \\
        Overtime status	& Yes \\
        Co\-worker relationship satisfaction & Very dissatisfied \\
        Job involvement & Disengaged \\
        Work environment satisfaction	& Very dissatisfied \\
        \midrule
        \textbf{PREDICTION:} & \textbf{RESIGN} \\
        \midrule
        \multicolumn{2}{p{7.8cm}}{Tanya's details were supplied to an automated system used by the Human Resources Department that predicted that Tanya would likely \textbf{resign} due to multiple factors, which includes dissatisfaction with her relationship with her co-workers, her low job involvement levels, and increased overtime. At present, Tanya does overtime, her co-worker relationship satisfaction is `very dissatisfied' and her job involvement level is `disengaged'. These ratings are due to an increase in business-related travel, which added overtime and took time away that Tanya usually spends with her colleagues.} \\
        \bottomrule
    \end{tabular}
    
    \label{tab:sample_employee_profile}
\end{table}

\subsection{Procedure}

We conducted our studies using Amazon MTurk~\cite{buhrmester2016amazon}. We designed and administered the experiments as a Qualtrics\footnote{\url{https://www.qualtrics.com/}} survey. Before the experiments, we received ethics approval from our institution. Participants were paid USD \$5 (\$15/hr)  for participating in the study. 

In total, 165 MTurkers completed the studies: 82 in Study 1 and 83 in Study 2. A power analysis conducted using G*Power~\cite{Faul2007-bo} suggested that the sample size required to achieve 80\% power for detecting a medium effect, at a significance criterion of $\alpha = 0.05$, was $N = 40$ per condition. We recruited \emph{Masters} workers\footnote{\url{https://www.mturk.com/worker/help}} from the United States. We first administered the plain language statement and then a consent form. Participants then answered one attention check question to filter out automated respondents and filled in the demographics questionnaire. Following this, we randomly allocated each participant to one of the two studies to ensure they participated in only one study. Figure~\ref{fig:study_procedure} presents the study procedure visually. Sample tasks from the two studies and a list of all scenarios and explanations are available in the appendix.

\begin{figure*}[t]
    \centering
    \includegraphics[width=\textwidth]{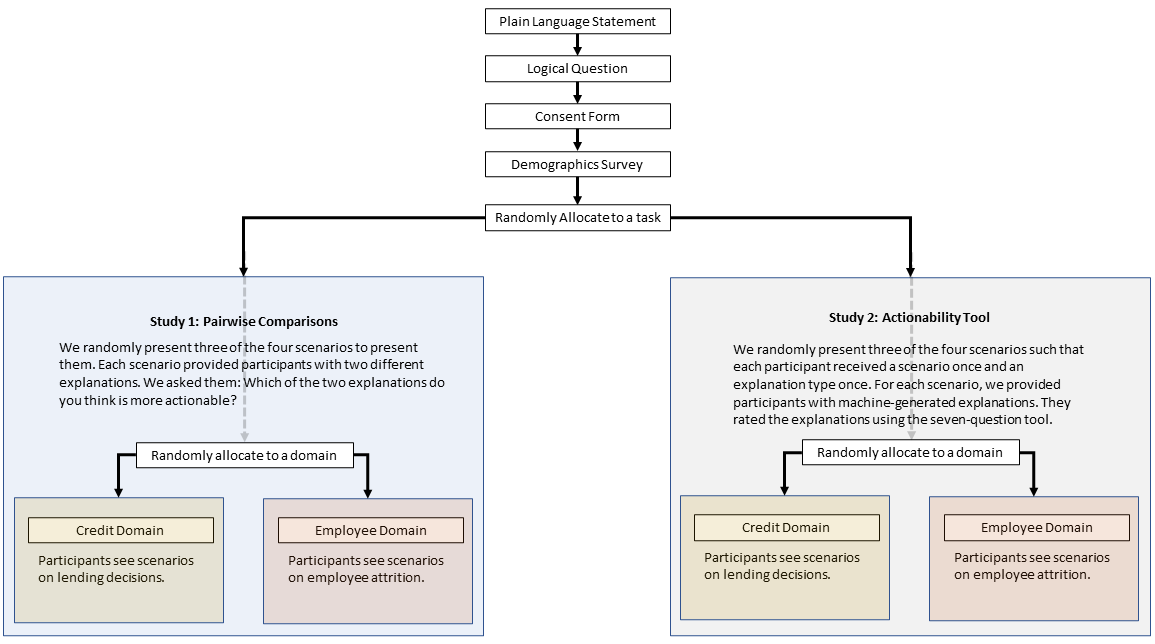}
    \caption{Study procedure: We run the two studies in parallel to allocate different sets of participants to the two studies.}
    \label{fig:study_procedure}
\end{figure*}

\subsection{Models and explanation generation}
\label{sec:study:ml}

For the credit scoring domain, we trained a logistic regression model to predict whether a borrower would default on a loan using the Lending Club dataset\footnote{\url{https://www.kaggle.com/husainsb/lendingclub-issued-loans\#lc_loan.csv}}. The model achieved an accuracy of 85\%. Similarly, for the employee satisfaction domain, we trained a logistic regression model to predict whether an employee would likely resign using an existing dataset\footnote{\url{https://www.kaggle.com/pavansubhasht/ibm-hr-analytics-attrition-dataset}}. The model achieved an accuracy of 76\%. 

We used the algorithm proposed by \citeN{Gurumoorthy2019-oe}, called ProtoDash, to generate one prototype per class. Their algorithm selects prototypes for a given sparsity level $m$
and associates non-negative weights with the prototypes to indicate their importance. We did not restrict the sparsity level and generated more than 10 prototypes for each class. We then selected the most important prototype, guided by the weights produced by the algorithm. We used the same prototype for each instance presented to the participants.

We used the algorithm proposed by~\citeN{Russell2019-jh} to generate the counterfactual explanations. This generates many diverse counterfactual explanations. For our study, we generated only one counterfactual, $\cf$, that is closest to the factual instance, $\x$, with a different outcome by solving the following problem:
\vspace{-1pt}
\begin{equation}
    \arg\min_{c}\max_{\tau} \left(\|x-c\| +  \tau\left(f\left(x\right)-f\left(c\right)\right)\right)
    \label{eq:russell}
\end{equation}

The distance function used by \citeN{Russell2019-jh} is $\ell_1$, weighted by the inverse Median Absolute Deviation ($\|.\|_{1,MAD}$). The function $\tau$ maximises the difference between the prediction of the counterfactual, $\cf$ and the factual point, $\x$. This means that the counterfactual instance we use in our studies is the closest point to the instance we explain with a different outcome. 

We used the method proposed by \citeN{Singh2023-is}  to generate directive (directive-specific) explanations. We first used  Russell's algorithm \cite{Russell2019-jh}  to generate one counterfactual instance, $\vec{c}$. The counterfactual instance $\vec{c}$ represents the goal state for the directive. We implemented a generic Monte-Carlo Tree Search algorithm \cite{Browne2012-mj} to find directives for that goal state. See the Appendix for the algorithms and parameters.

Given the respective prototype, counterfactual, and directive explanations, we used a simple template to create natural language and tabular-based explanations. All of the explanations for the three types of scenarios can be found in the appendix.

We acknowledge that these three approaches do not cover the entire range of algorithms, for example-based explainability methods. However, we remind the reader that the purpose of this study is to evaluate the actionability tool, not to evaluate the actionability of different explainability methods.

\subsection{Study 1: Pairwise comparisons}
In Study 1, we were interested in whether the participants could differentiate between the explanation types regarding their actionability. We did not define `actionability' for the participants, as any definition we provided would naturally align with our conception of actionability, which was a direct input to the design of the actionability tool. Offering participants a definition would introduce a clear experimenter bias.

Our experiment measured \textit{perceived actionability}, whereas to measure ``actual actionability", i.e. which explanations people \textit{use} to in realistic decision scenarios to change the outcome, is a non-trivial task that is only worth doing if we have a tool that has been empirically demonstrated as promising. Therefore, we chose to use perceived actionability as a validation for Study 2.

We randomly allocated participants to one domain so that they received scenarios from only one domain, and we randomly selected three of the four scenarios to present them. After each scenario, we provided participants with two different explanations; for example, a counterfactual and a prototypical one. We asked the participants the following question, and they responded by selecting one of the two explanations:

\begin{quote}
    \textit{The system has generated the following two explanations for you. Which of the two explanations do you think is more actionable?}
\end{quote}

\subsection{Study 2: Rating explanations via the Actionability Tool}
For participants allocated to Study 2, we randomly assigned them to a domain so that they received scenarios only from that domain; participants in this study saw lending decision scenarios or employee turnover scenarios, but not both. For each domain, we randomly selected three of the four scenarios, and each participant received a scenario once and an explanation type once. We evaluated the actionability of three explanation types. We present a scenario once and get the participant to rate each explanation type once. The scenario and explanation type order was randomised to reduce ordering effects.  

In the credit scoring domain, we asked participants to imagine they were loan applicants, and the information (explanation) was to help them change the model's decision (e.g. get their loan approved). In the employee turnover domain, we asked participants to imagine that they were the concerned employee's supervisor, and the information was to help them prevent employees from leaving the organisation.

We provided participants with machine-generated explanations for each scenario, using the approaches discussed in Section~\ref{sec:study:ml}. They rated the explanations using the seven-question tool. We presented the questions in the same order each time and in the order discussed in the paper (Q1 to Q7). 

\section{Results}
When required, we performed a non-parametric Friedman test for both studies and Nemenyi posthoc tests. We used non-parametric tests, such as the Shapiro-Wilk test, which showed that the data was not normal. Across both studies, the mean task completion time was 26 minutes $\small(SD=12~mins)$. 

\subsection{Participant Demographics}

Across the studies, around 49\% self-identified as men, 47\% as women, and 4\% did not state their gender. Age: 24\% were 25-34, 33\% were 35-44, 22\% were 45-54, and the rest were above 55 (21\%). Education: 12\% were High school graduates, 21\% had some college education but no degree, 50\% had an Associate or Bachelor's degree, 11\% had a Master's degree, 2\% had a Doctoral degree, and 4\% had a Professional degree or did not state their education level.

We evaluated all results below to establish if there is any correlation to gender, age, and education, and found no significant difference between these groups; therefore, these results are omitted.

\subsection{Study 1: Pairwise comparisons}

\begin{figure}
    \centering
    \includegraphics[scale=0.9]{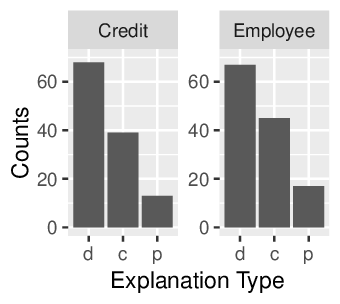}
    \caption{Results of pairwise comparison between explanation types. The y-axis is the number of times each explanation type was perceived to be more actionable (d = directive; c = counterfactual; and p = prototype).}
    \label{fig:pairwise_preferences}
\end{figure}

A key intent of this study was to provide data to compare against the results of Study 2. We had 83 participants respond (40 in credit scoring and 43 in employee turnover). Figure~\ref{fig:pairwise_preferences} shows the results.

Overall, we collected 249 rankings. A chi-square goodness-of-fit test shows that the likelihood of observing the data if the choices for the most actionable explanations were random was low in the credit domain, $\chi^2 \left(2, N = 119\right) = 37.9, p < 0.001$, and the employee domain $\chi^2 \left(2, N = 128\right) = 29.2, p < 0.001$. For each participant, we counted the number of times they selected each explanation type and used a Friedman test to identify if there was a significant difference in the number of times each explanation type was perceived to be actionable. Results of the Friedman test (Table~\ref{tab:friedman_test_pairwise}) confirm that the directives were perceived to be more actionable than counterfactuals, which were perceived to be more actionable than prototypical explanations. These results confirm our hypothesis that directive explanations were perceived as more actionable than counterfactual ones, which were more actionable than prototypical explanations.

\begin{table}[ht!]
    \small
    \centering
    \caption{The table shows the Friedman and Nemenyi post-hoc test results for pairwise comparisons (Study 1). }
    \label{tab:friedman_test_pairwise}
    \begin{tabular}{ccc}
    \toprule
         & Credit & Employee  \\
         \midrule
         $\chi^2\left(2\right)$ & $40.9$ & $31.4$   \\
           $p$ & $\textbf{\texttt{<}0.001}$ & $\textbf{\texttt{<}0.001}$   \\
           $Kendall's W$ & $0.51$ & $0.37$   \\
          c vs d & $\textbf{0.003}$ & $\textbf{0.04}$   \\
          c vs p & $\textbf{0.01}$ & $\textbf{0.007}$ \\
          d vs p & $\textbf{\texttt{<}0.001}$ & $\textbf{\texttt{<}0.001}$ \\

          \bottomrule
    \end{tabular}
    
\end{table}

\begin{figure*}[ht!]
    \centering
    \includegraphics[width=\textwidth]{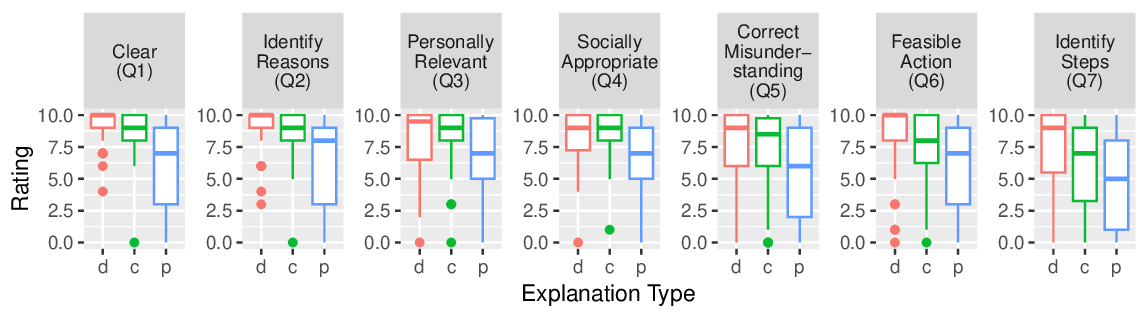}
    \caption{Ratings for explanations in the credit domain. Explanation types: d = directive; c = counterfactual; p = prototype.}
    \label{fig:rating_credit}
\end{figure*}

\begin{figure*}[ht!]
    \centering
    \includegraphics[width=\textwidth]{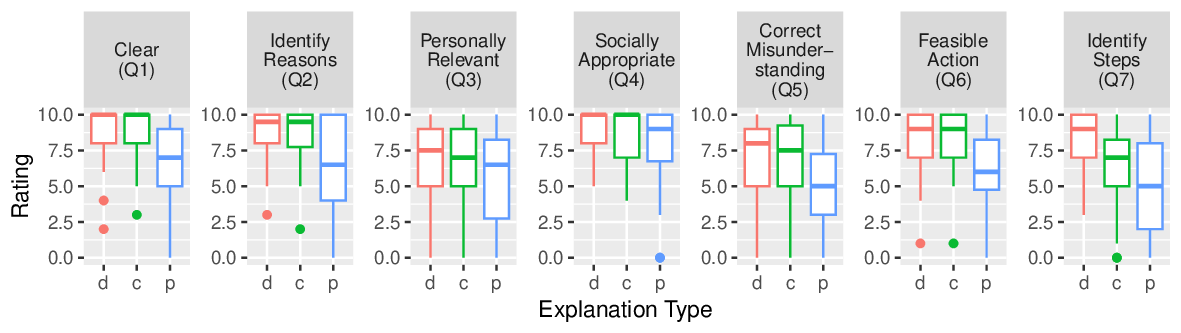}
    \caption{Ratings for the explanation types in the employee turnover domain. Explanation types: d = directive; c = counterfactual; p = prototype.}
    \label{fig:rating_hr}
\end{figure*}

\begin{table*}
    \small
    \centering
    \caption{The results of the Friedman and Nemenyi post-hoc tests for the two domains. For each domain, the first three rows are the Friedman test statistics, and the last three are $p$ values from the post-hoc tests comparing the rating between explanation types. Significant values are shown in bold.  [c, d, p for counterfactual, directive, prototypical explanations, respectively]}
    \label{tab:friedman_test_ratings}
    \begin{tabular}{p{1.2cm}p{1.5cm}p{1.2cm}p{1.2cm}p{1.5cm}p{1.6cm}p{1.5cm}p{1.2cm}p{1.2cm}}
    \toprule
          &  
          & 
          Clear (Q1) & 
          Identify Reasons (Q2) & 
          Personally Relevant (Q3) & 
          Socially Appropriate (Q4) & 
          Correct Misunderstanding (Q5) & 
          Feasible Action (Q6) & 
          Identify Steps (Q7) \\
         \midrule
          \multirow{6}{*}{Credit} & 
          $\chi^2\left(2\right)$ & 
          $31.9$ & 
          $19.7$  & 
          $15.5$ & 
          $15.6$ & 
          $14.7$ & 
          $16.5$ & 
          $31.5$ \\
          
           & 
           $p$ & 
           $\textbf{\texttt{<}0.001}$ & 
           $\textbf{\texttt{<}0.001}$  & 
           $\textbf{\texttt{<}0.001}$ & 
           $\textbf{\texttt{<}0.001}$ & 
           $\textbf{\texttt{<}0.001}$ & 
           $\textbf{\texttt{<}0.001}$ & 
           $\textbf{\texttt{<}0.001}$ \\
           
           & 
           $Kendall's W$ & 
           $0.38$ & 
           $0.33$  & 
           $0.19$ & 
           $0.19$ & 
           $0.18$ & 
           $0.20$ & 
           $0.38$ \\
          
          & 
          c vs d & 
          $0.82$ & 
          $\textbf{\texttt{<}0.001}$  & 
          $0.94$ & 
          $0.88$ & 
          $0.55$ & 
          $0.30$ & 
          $\textbf{0.008}$ \\
          
          & 
          c vs p & 
          $\textbf{0.003}$ & 
          $\textbf{\texttt{<}0.001}$ & 
          $\textbf{0.02}$ & 
          $\textbf{0.009}$ & 
          $0.07$ & 
          $0.12$ & 
          $0.15$ \\
          
          & 
          d vs p & 
          $\textbf{\texttt{<}0.001}$ & 
          $\textbf{0.002}$ & 
          $\textbf{0.009}$ & 
          $\textbf{0.04}$ & 
          $\textbf{0.004}$ & 
          $\textbf{0.002}$ & 
          $\textbf{\texttt{<}0.001}$ \\
          \midrule
          
          \multirow{6}{*}{Employee} & 
          $\chi^2\left(2\right)$ & 
          $17.3$ & 
          $18.1$  & 
          $4.66$ & 
          $1.98$ & 
          $13.7$ & 
          $25.1$ & 
          $34.6$ \\
          
          & 
          $p$ & 
          $\textbf{\texttt{<}0.001}$ & 
          $\textbf{\texttt{<}0.001}$  & 
          $0.10$ & 
          $0.38$ & 
          $\textbf{0.001}$ & 
          $\textbf{\texttt{<}0.001}$ & 
          $\textbf{\texttt{<}0.001}$ \\
          
           & 
          $Kendall's W$ & 
          $0.22$ & 
          $0.23$  & 
          $0.06$ & 
          $0.02$ & 
          $0.17$ & 
          $0.31$ & 
          $0.43$ \\
          
          & 
          c vs d & 
          $0.87$ & 
          $0.96$  & 
          $0.84$ & 
          $0.97$ & 
          $0.96$ & 
          $0.96$ & 
          $\textbf{0.003}$ \\
          & 
          c vs p & 
          $\textbf{0.006}$ & 
          $\textbf{0.006}$ & 
          $0.44$ & 
          $0.61$ & 
          $\textbf{0.01}$ & 
          $\textbf{0.002}$ & 
          $0.14$ \\
          
          & 
          d vs p & 
          $\textbf{0.02}$ & 
          $\textbf{0.01}$ & 
          $0.17$ & 
          $0.75$ & 
          $\textbf{0.02}$ & 
          $\textbf{\texttt{<}0.001}$ & 
          $\textbf{\texttt{<}0.001}$ \\
          \bottomrule
    \end{tabular}
    
\end{table*}

\subsection{Study 2: Rating explanations via the Actionability Tool}
A total of 82 participants responded (42 in credit scoring and 40 in employee turnover). We analyse each question separately rather than combining responses into a holistic score. This approach acknowledges the diversity in constructs, thus ensuring insights are not lost in aggregation. Finally, this method prevents high scores in one area might mask low scores in another, thus ensuring a balanced and accurate representation of the subject matter, which is crucial as we move towards an effective evaluation tool.

\subsubsection{Credit Scoring}
We plot the median ratings for the seven questions in Figure~\ref{fig:rating_credit} and show the results of the Friedman and Nemenyi post-hoc tests in Table~\ref{tab:friedman_test_ratings}.

Analysis reveals no significant difference in perceived actionability between counterfactual and directive explanations for five of the seven questions (Q1, Q3 - Q6). Directive explanations were better at breaking down actions into explicit steps (Q7) and clarifying decision reasons (Q2) than counterfactual and prototypical explanations across the board.

Interestingly, despite our expectations for directive explanations to facilitate the identification of a feasible action (Q6), counterfactual explanations proved equally effective, often offering implicit actionable insights recognised by participants. One participant observed that counterfactual explanations might not detail specific actions, but they assumed an average person's awareness of potential credit score improvements.

\subsubsection{Employee Turnover}
We plot the median ratings in Figure~\ref{fig:rating_hr} and show the Friedman and Nemenyi post-hoc test results in Table~\ref{tab:friedman_test_ratings}. In the employee domain, directive and counterfactual explanations were preferred over prototypical ones across four questions (Q1, Q2, Q5, Q6). There was no significant difference between counterfactual and directive explanations for six out of seven items. Directive explanations were valued for their clarity in identifying actionable steps (Q7), mirroring findings from the credit scoring domain.

We interpret the mixed results between counterfactual and prototypical explanations and between directive and prototypical ones as influenced by the framing of the participants' task in the employee domain, where explanations were addressed not to participants but to the hypothetical employees they supervised. Any suggested action would have been for the concerned employee, not the participant (employee's supervisor). Therefore, the explanation was not \textit{personally relevant (Q3)} to the supervisor but to the employee. This shift in relevance might explain why counterfactual explanations were slightly favoured for specific questions, suggesting adequacy in indicating necessary changes without direct instruction.

Across both domains, the consistent differentiation in ratings between prototypical and directive explanations supports the effectiveness of the evaluation tool, particularly in distinguishing between these explanation types' actionability.

\section{Discussion}
We proposed a tool to judge the actionability of information to enable algorithmic recourse. We devised the seven questions using existing instruments. We chose three example-based explanation methods that we believed would have differing levels of actionability: prototypical, counterfactual, and directive explanations, from lowest to highest. The results of our first study validated our choices: participants rated example explanations according to our intuitions, effectively giving us a `ground truth' about the actionability of three types of explanations. The results of our second study further showed that participants using the actionability tool gave ratings that correlated with findings from the first step.

\textbf{The tool effectively distinguished between the actionability of directive, counterfactual, and prototypical explanations across both domains:}~One of the tool's strengths is its ability to differentiate between the perceived actionability of different explanations. Directive explanations were consistently rated as more actionable across both domains, especially against prototypical explanations, highlighting the tool's effectiveness in capturing participants' preferences for explanations that provide clear, direct guidance. In Study 1, directive explanations were also perceived most actionable. 

Directive explanations consistently appeared most actionable, particularly in providing explicit steps (Q1). This trend emphasises the importance of clarity and specificity in enhancing the actionability of explanations. Study 1 shows directive explanations were more actionable than counterfactual explanations, and the higher rating of directive explanations in Study 2 explains the important factor of information being clear and easy to understand (Q1) for this differentiation. The differentiation between directive and counterfactual explanations highlights the tool's utility in assessing subtle aspects of the actionability of explanations.

Further, the ability of the questions Identify Reasons (Q2), Correct Misunderstanding (Q5), and Feasible Action (Q6) to differentiate prototypical from counterfactual and directive explanations further supports the tool's effectiveness.

\textbf{Contextual factors influence explanation ratings:} The results demonstrate the importance of universally valued factors like clarity and actionable guidance across credit scoring and employee turnover domains. However, it highlighted how contextual factors, particularly participant roles, can affect explanation ratings. In the credit domain, participants directly related to explanations as loan applicants, making items like Personal Relevance (Q3) more impactful. Conversely, in the employee domain, where participants acted in a supervisory role, the importance of factors such as Social Appropriateness (Q4) increased due to the ability to guide supervisors in determining interventions. This advisory perspective made Personally Relevant (Q3) and Correct Misunderstanding (Q5) less important due to their focus on another's career path, showcasing how the perceived actionability of explanations varies with context and the recipient's role. 

\subsection{Limitations and future works}
The settings in the domains were different, so a direct comparison of the ratings in the two domains is not appropriate. This highlights the need to investigate the tool or questions in different domains and settings, with more participants and explanation types. Also, we recognise that users with similar profiles as represented in the algorithmic decision making may have different action preferences, so other methods would need to be incorporated to accommodate personalisation \cite{Singh_Manan2023}, or account for social factors \cite{Singh2023-is}. While we used just one method each for prototype, counterfactual, and directive explanations, others exist; e.g.\ \cite{Verma2020-ud} for generating counterfactuals. Future work could explore how actionability changes with different counterfactual methods. Still, we again note that the purpose of the current study is to evaluate the actionability tool, not specific algorithms or types of counterfactuals.

Our tool assesses \emph{perceived} actionability, and we do not investigate how this relates to \emph{actual} actionability -- that is, the ability for people to understand and implement actionable approaches to recourse. Assessing perceived actionability is one important step, but additional studies are required to determine how well this correlates with behavioural actionability. Future work should examine the capacity of various recourse algorithms, including causal recourse models, to generate actionable outcomes, yielding data to validate the tool further. 

\section{Conclusion}

Our contribution demonstrates the feasibility of creating a meaningful human-centred actionability tool to assess the actionability of information provided to enable algorithmic recourse.
The tool is based on existing actionability instruments. Evidence for the usefulness of the tool is that the results of our two studies align: the ratings allocated by participants using the assessment tool align with the pairwise judgements provided by independent participants. This supports our hypothesis that the tool is useful in discriminating explanation actionability between different methods. However,  the tool may need to be adapted for specific domains. 
Further validation requires input from other stakeholders in both domains, experimentation with other methods for algorithmic recourse and in other domains.
The longer-term goal of this work is to enable researchers and practitioners to use the questionnaire as a design tool to help produce more actionable information.
\subsection{Acknowledgements}
This project is supported by Australian Research Council (ARC) Discovery Grant DP190103414: \emph{Explanation in Artificial Intelligence: A Human-Centred Approach}.

\bibliographystyle{IEEEtranN}
\bibliography{references}

\begin{thebibliography}{40}
\providecommand{\natexlab}[1]{#1}
\providecommand{\url}[1]{#1}
\csname url@samestyle\endcsname
\providecommand{\newblock}{\relax}
\providecommand{\bibinfo}[2]{#2}
\providecommand{\BIBentrySTDinterwordspacing}{\spaceskip=0pt\relax}
\providecommand{\BIBentryALTinterwordstretchfactor}{4}
\providecommand{\BIBentryALTinterwordspacing}{\spaceskip=\fontdimen2\font plus
\BIBentryALTinterwordstretchfactor\fontdimen3\font minus \fontdimen4\font\relax}
\providecommand{\BIBforeignlanguage}[2]{{%
\expandafter\ifx\csname l@#1\endcsname\relax
\typeout{** WARNING: IEEEtranN.bst: No hyphenation pattern has been}%
\typeout{** loaded for the language `#1'. Using the pattern for}%
\typeout{** the default language instead.}%
\else
\language=\csname l@#1\endcsname
\fi
#2}}
\providecommand{\BIBdecl}{\relax}
\BIBdecl

\bibitem[Karimi et~al.(2022)Karimi, Barthe, Sch\"{o}lkopf, and Valera]{karimi2022}
\BIBentryALTinterwordspacing
A.-H. Karimi, G.~Barthe, B.~Sch\"{o}lkopf, and I.~Valera, ``A survey of algorithmic recourse: Contrastive explanations and consequential recommendations,'' \emph{ACM Comput. Surv.}, vol.~55, no.~5, Dec 2022. [Online]. Available: \url{https://doi.org/10.1145/3527848}
\BIBentrySTDinterwordspacing

\bibitem[Sullivan and Verreault{-}Julien(forthcoming)]{Sullivan2022-sm}
E.~Sullivan and P.~Verreault{-}Julien, ``From explanation to recommendation: Ethical standards for algorithmic recourse,'' \emph{Proceedings of the 2022 AAAI/ACM Conference on AI, Ethics, and Society (AIES '22)}, forthcoming.

\bibitem[Rasouli and Chieh~Yu(2022)]{rasouli2022}
P.~Rasouli and I.~Chieh~Yu, ``Care: Coherent actionable recourse based on sound counterfactual explanations,'' \emph{International Journal of Data Science and Analytics}, pp. 1--26, 2022.

\bibitem[Kelechi and Jiao(2023)]{Kelechi2023-aw}
N.~Kelechi and L.~Jiao, ``Quantifying actionability: Evaluating {Human-Recipient} models,'' \emph{IEEE Access}, vol.~11, pp. 119\,811--119\,823, 2023.

\bibitem[Ustun et~al.(2019)Ustun, Spangher, and Liu]{Ustun2019-ic}
B.~Ustun, A.~Spangher, and Y.~Liu, ``Actionable recourse in linear classification,'' in \emph{Proceedings of the Conference on Fairness, Accountability, and Transparency}, ser. FAT* '19.\hskip 1em plus 0.5em minus 0.4em\relax New York, NY, USA: Association for Computing Machinery, Jan. 2019, pp. 10--19.

\bibitem[Leavitt and Morcos(2020)]{Leavitt2020-kj}
\BIBentryALTinterwordspacing
M.~L. Leavitt and A.~Morcos, ``{Towards falsifiable interpretability research},'' Oct. 2020. [Online]. Available: \url{http://arxiv.org/abs/2010.12016}
\BIBentrySTDinterwordspacing

\bibitem[Miller et~al.(2017)Miller, Howe, and Sonenberg]{Miller2017-cu}
\BIBentryALTinterwordspacing
T.~Miller, P.~Howe, and L.~Sonenberg, ``{Explainable {AI}: Beware of Inmates Running the Asylum Or: How I Learnt to Stop Worrying and Love the Social and Behavioural Sciences},'' \emph{Proceedings of the Explainable AI Workshop @ IJCAI 2017}, Dec. 2017. [Online]. Available: \url{http://arxiv.org/abs/1712.00547}
\BIBentrySTDinterwordspacing

\bibitem[Shoemaker et~al.(2014)Shoemaker, Wolf, and Brach]{Shoemaker2014-wo}
S.~J. Shoemaker, M.~S. Wolf, and C.~Brach, ``\BIBforeignlanguage{en}{Development of the patient education materials assessment tool ({PEMAT)}: a new measure of understandability and actionability for print and audiovisual patient information},'' \emph{\BIBforeignlanguage{en}{Patient Educ. Couns.}}, vol.~96, no.~3, pp. 395--403, Sep. 2014.

\bibitem[Byrne(2023)]{Byrne-ijcai2023survey}
R.~M. Byrne, ``Good explanations in explainable artificial intelligence (xai): Evidence from human explanatory reasoning,'' in \emph{Proceedings of the Thirty-Second International Joint Conference on Artificial Intelligence, {IJCAI-23}}, E.~Elkind, Ed., 2023, pp. 6536--6544, survey Track.

\bibitem[Hoffman et~al.(2023)Hoffman, Mueller, Klein, and Litman]{Hoffman2023a}
R.~R. Hoffman, S.~T. Mueller, G.~Klein, and J.~Litman, ``Measures for explainable ai: Explanation goodness, user satisfaction, mental models, curiosity, trust, and human-ai performance,'' \emph{Frontiers in Computer Science}, vol.~5, 2023.

\bibitem[Molnar(2020)]{Molnar2020-fj}
C.~Molnar, \emph{\BIBforeignlanguage{en}{Interpretable Machine Learning}}.\hskip 1em plus 0.5em minus 0.4em\relax Lulu.com, 2020.

\bibitem[Gurumoorthy et~al.(2019)Gurumoorthy, Dhurandhar, Cecchi, and Aggarwal]{Gurumoorthy2019-oe}
K.~S. Gurumoorthy, A.~Dhurandhar, G.~Cecchi, and C.~Aggarwal, ``Efficient data representation by selecting prototypes with importance weights,'' in \emph{2019 {IEEE} International Conference on Data Mining ({ICDM})}.\hskip 1em plus 0.5em minus 0.4em\relax ieeexplore.ieee.org, Nov. 2019, pp. 260--269.

\bibitem[Wachter et~al.(2017)Wachter, Mittelstadt, and Russell]{Wachter2017-fv}
S.~Wachter, B.~Mittelstadt, and C.~Russell, ``Counterfactual explanations without opening the black box: Automated decisions and the {GDPR},'' \emph{Harvard Journal of Law \& Technology}, vol.~31, no.~2, pp. 841--887, Oct. 2017.

\bibitem[Russell(2019)]{Russell2019-jh}
C.~Russell, ``Efficient search for diverse coherent explanations,'' in \emph{Proceedings of the Conference on Fairness, Accountability, and Transparency}, ser. FAT* '19.\hskip 1em plus 0.5em minus 0.4em\relax New York, NY, USA: Association for Computing Machinery, Jan. 2019, pp. 20--28.

\bibitem[Singh et~al.(2023{\natexlab{a}})Singh, Miller, Lyons, Sonenberg, Velloso, Vetere, Howe, and Dourish]{Singh2023-is}
R.~Singh, T.~Miller, H.~Lyons, L.~Sonenberg, E.~Velloso, F.~Vetere, P.~Howe, and P.~Dourish, ``\BIBforeignlanguage{en}{Directive explanations for actionable explainability in machine learning applications},'' \emph{\BIBforeignlanguage{en}{ACM Trans. Interact. Intell. Syst.}}, Jan. 2023.

\bibitem[Mothilal et~al.(2020)Mothilal, Sharma, and Tan]{Mothilal2020-qi}
R.~K. Mothilal, A.~Sharma, and C.~Tan, ``Explaining machine learning classifiers through diverse counterfactual explanations,'' in \emph{Proceedings of the 2020 Conference on Fairness, Accountability, and Transparency}, ser. FAT* '20.\hskip 1em plus 0.5em minus 0.4em\relax New York, NY, USA: Association for Computing Machinery, Jan. 2020, pp. 607--617.

\bibitem[Aryal and Keane(2023)]{Aryal-Keane-ijcai2023survey}
S.~Aryal and M.~T. Keane, ``Even if explanations: Prior work, desiderata \& benchmarks for semi-factual xai,'' in \emph{Proceedings of the Thirty-Second International Joint Conference on Artificial Intelligence, {IJCAI-23}}, E.~Elkind, Ed., 2023, pp. 6526--6535, survey Track.

\bibitem[Leofante and Lomuscio(2023)]{Leofante2023-dt}
F.~Leofante and A.~Lomuscio, ``Towards robust contrastive explanations for {Human-Neural} multi-agent systems,'' in \emph{Proceedings of the 2023 International Conference on Autonomous Agents and Multiagent Systems}, 2023, pp. 2343--2345.

\bibitem[Gajcin(2023)]{Gajcin2023-eg}
J.~Gajcin, ``Counterfactual explanations for reinforcement learning agents,'' in \emph{Proceedings of the 2023 International Conference on Autonomous Agents and Multiagent Systems}.\hskip 1em plus 0.5em minus 0.4em\relax southampton.ac.uk, 2023, pp. 2925--2927.

\bibitem[Poyiadzi et~al.(2020)Poyiadzi, Sokol, Santos-Rodriguez, De~Bie, and Flach]{Poyiadzi2020-tp}
R.~Poyiadzi, K.~Sokol, R.~Santos-Rodriguez, T.~De~Bie, and P.~Flach, ``{FACE}: Feasible and actionable counterfactual explanations,'' in \emph{Proceedings of the {AAAI/ACM} Conference on {AI}, Ethics, and Society}, ser. AIES '20.\hskip 1em plus 0.5em minus 0.4em\relax New York, NY, USA: Association for Computing Machinery, Feb. 2020, pp. 344--350.

\bibitem[Bhattacharya et~al.(2023)Bhattacharya, Ooge, Stiglic, and Verbert]{bhattacharya2023directive}
A.~Bhattacharya, J.~Ooge, G.~Stiglic, and K.~Verbert, ``Directive explanations for monitoring the risk of diabetes onset: Introducing directive data-centric explanations and combinations to support what-if explorations,'' in \emph{Proceedings of the 28th International Conference on Intelligent User Interfaces}, 2023, pp. 204--219.

\bibitem[Karimi et~al.(2021)Karimi, Sch{\"o}lkopf, and Valera]{Karimi2021-ke}
A.-H. Karimi, B.~Sch{\"o}lkopf, and I.~Valera, ``Algorithmic recourse: from counterfactual explanations to interventions,'' in \emph{Proceedings of the 2021 {ACM} Conference on Fairness, Accountability, and Transparency}, ser. FAccT '21.\hskip 1em plus 0.5em minus 0.4em\relax New York, NY, USA: Association for Computing Machinery, Mar. 2021, pp. 353--362.

\bibitem[Wang et~al.(2002)Wang, Zhou, and Han]{Wang2002-ys}
K.~Wang, S.~Zhou, and J.~Han, ``Profit mining: From patterns to actions,'' in \emph{Advances in Database Technology --- {EDBT} 2002}.\hskip 1em plus 0.5em minus 0.4em\relax Springer Berlin Heidelberg, 2002, pp. 70--87.

\bibitem[Tzacheva and Ranganathan(2020)]{Tzacheva2020-kt}
A.~A. Tzacheva and J.~Ranganathan, ``\BIBforeignlanguage{en}{Emotion mining from text for actionable recommendations detailed survey},'' \emph{\BIBforeignlanguage{en}{Int. J. Data Min. Model. Manag.}}, vol.~12, no.~2, p. 143, 2020.

\bibitem[{Redmiles} et~al.(2020){Redmiles}, {Warford}, {Jayanti}, {Koneru}, and {others}]{Redmiles2020-jb}
{Redmiles}, {Warford}, {Jayanti}, {Koneru}, and {others}, ``A comprehensive quality evaluation of security and privacy advice on the web,'' \emph{29th USENIX Security}, 2020.

\bibitem[HakemZadeh and Baba(2016)]{HakemZadeh2016-ba}
F.~HakemZadeh and V.~V. Baba, ``Measuring the actionability of evidence for evidence-based management,'' \emph{Management Decision}, vol.~54, no.~5, pp. 1183--1204, Jan. 2016.

\bibitem[Syed et~al.(2010)Syed, Mingers, and Murray]{Syed2010-bf}
J.~Syed, J.~Mingers, and P.~A. Murray, ``Beyond rigour and relevance: A critical realist approach to business education,'' \emph{Management Learning}, vol.~41, no.~1, pp. 71--85, Feb. 2010.

\bibitem[{Converse} et~al.(1993){Converse}, {Cannon-Bowers}, and {others}]{Converse1993-uo}
{Converse}, {Cannon-Bowers}, and {others}, ``Shared mental models in expert team decision making,'' \emph{Individual and group decision making}, pp. 221--246, 1993.

\bibitem[{Scholl} et~al.(2011){Scholl}, {Koelewijn-van Loon}, {Sepucha}, and {others}]{Scholl2011-gp}
{Scholl}, {Koelewijn-van Loon}, {Sepucha}, and {others}, ``Measurement of shared decision making--a review of instruments,'' \emph{Z. Evid. Fortbild. Qual. Gesundhwes.}, vol. 105, no.~4, pp. 313--324, 2011.

\bibitem[Sainfort and Booske(2000)]{Sainfort2000-ta}
F.~Sainfort and B.~C. Booske, ``\BIBforeignlanguage{en}{Measuring post-decision satisfaction},'' \emph{\BIBforeignlanguage{en}{Med. Decis. Making}}, vol.~20, no.~1, pp. 51--61, Jan. 2000.

\bibitem[O'Connor(1995)]{OConnor1995-yb}
A.~O'Connor, ``Decision self-efficacy scale,'' \url{https://decisionaid.ohri.ca/docs/develop/user_manuals/UM_decision_selfefficacy.pdf}, 1995, accessed: 2023-1-18.

\bibitem[Rogers and Prentice-Dunn(1997)]{Rogers1997-pv}
R.~W. Rogers and S.~Prentice-Dunn, ``Protection motivation theory,'' \emph{Handbook of health behavior research 1: Personal and social determinants.}, vol.~1, no. 505, pp. 113--132, 1997.

\bibitem[Cranor(2008)]{Cranor2008-bl}
L.~F. Cranor, ``A framework for reasoning about the human in the loop,'' \url{https://www.usenix.org/legacy/event/upsec/tech/full_papers/cranor/cranor.pdf}, 2008, accessed: 2022-6-27.

\bibitem[Redmiles et~al.(2018)Redmiles, Zhu, Kross, Kuchhal, Dumitras, and Mazurek]{Redmiles2018-kq}
E.~M. Redmiles, Z.~Zhu, S.~Kross, D.~Kuchhal, T.~Dumitras, and M.~L. Mazurek, ``Asking for a friend: Evaluating response biases in security user studies,'' in \emph{Proceedings of the 2018 {ACM} {SIGSAC} Conference on Computer and Communications Security}, ser. CCS '18.\hskip 1em plus 0.5em minus 0.4em\relax New York, NY, USA: Association for Computing Machinery, Oct. 2018, pp. 1238--1255.

\bibitem[Venkatasubramanian and Alfano(2020)]{Venkatasubramanian2020-vz}
S.~Venkatasubramanian and M.~Alfano, ``The philosophical basis of algorithmic recourse,'' in \emph{Proceedings of the 2020 Conference on Fairness, Accountability, and Transparency}, ser. FAT* '20.\hskip 1em plus 0.5em minus 0.4em\relax New York, NY, USA: Association for Computing Machinery, Jan. 2020, pp. 284--293.

\bibitem[Buhrmester et~al.(2011)Buhrmester, Kwang, and Gosling]{buhrmester2016amazon}
M.~Buhrmester, T.~Kwang, and S.~D. Gosling, ``\BIBforeignlanguage{en}{Amazon's mechanical turk: A new source of inexpensive, yet high-quality, data?}'' \emph{\BIBforeignlanguage{en}{Perspect. Psychol. Sci.}}, vol.~6, no.~1, pp. 3--5, Jan. 2011.

\bibitem[Faul et~al.(2007)Faul, Erdfelder, Lang, and Buchner]{Faul2007-bo}
F.~Faul, E.~Erdfelder, A.-G. Lang, and A.~Buchner, ``\BIBforeignlanguage{en}{{G*Power} 3: a flexible statistical power analysis program for the social, behavioral, and biomedical sciences},'' \emph{\BIBforeignlanguage{en}{Behav. Res. Methods}}, vol.~39, no.~2, pp. 175--191, May 2007.

\bibitem[Browne et~al.(2012)Browne, Powley, Whitehouse, Lucas, Cowling, Rohlfshagen, Tavener, Perez, Samothrakis, and Colton]{Browne2012-mj}
C.~B. Browne, E.~Powley, D.~Whitehouse, S.~M. Lucas, P.~I. Cowling, P.~Rohlfshagen, S.~Tavener, D.~Perez, S.~Samothrakis, and S.~Colton, ``A survey of monte carlo tree search methods,'' \emph{IEEE Trans. Comput. Intell. AI Games}, vol.~4, no.~1, pp. 1--43, Mar. 2012.

\bibitem[Singh et~al.(2023{\natexlab{b}})Singh, Kancheti, Gupta, Ghalme, Jain, and C.~Krishnan]{Singh_Manan2023}
M.~Singh, S.~S. Kancheti, S.~Gupta, G.~Ghalme, S.~Jain, and N.~C.~Krishnan, ``Algorithmic recourse based on user’s feature-order preference,'' in \emph{Proceedings of the 6th Joint International Conference on Data Science \& Management of Data (10th ACM IKDD CODS and 28th COMAD)}, ser. CODS-COMAD '23.\hskip 1em plus 0.5em minus 0.4em\relax Association for Computing Machinery, 2023, p. 293–294.

\bibitem[Verma et~al.(2020)Verma, Dickerson, and Hines]{Verma2020-ud}
S.~Verma, J.~Dickerson, and K.~Hines, ``Counterfactual explanations for machine learning: A review,'' \emph{ArXiv e-prints}, Oct. 2020.

\end{thebibliography}

\section*{Biography}
\vspace{-40pt}
\begin{IEEEbiography}[{\includegraphics[width=0.8in,height=1.25in,clip,keepaspectratio]{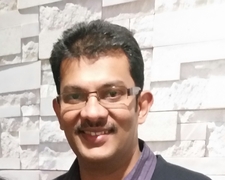}}]{Ronal Singh} Ronal is a Research Scientist at CSIRO's Data61 and CINTEL FSP, focusing on AI-assisted decision-making, and explainable AI. He previously held positions at the University of Melbourne, where he also earned his PhD in 2018, and holds BSc and MSc degrees in Computer Science from the University of the South Pacific, Fiji.
\end{IEEEbiography}
\vspace{-60pt}
\begin{IEEEbiography}[{\includegraphics[width=0.7in,height=1.2in,clip,keepaspectratio]{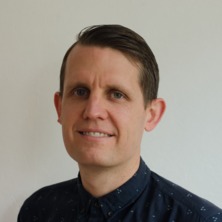}}]{Tim Miller} Tim is a Professor of AI at The University of Queensland, focusing on using machine learning, AI planning, and cognitive science for improved decision-making, including work in explainable AI and human-AI collaboration. Before this, he was a Professor at The University of Melbourne, and remains an honorary professor there.
\end{IEEEbiography}
\vspace{-60pt}
\begin{IEEEbiography}[{\includegraphics[width=0.75in,height=1.25in,clip,keepaspectratio]{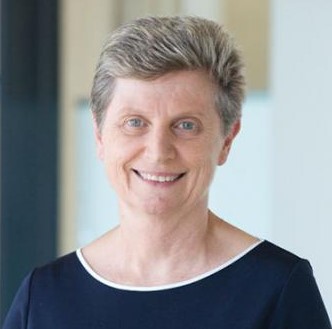}}]{Liz Sonenberg} Liz is the Pro Vice Chancellor of Systems Innovation and Professor at the University of Melbourne, focusing on human-centred AI for decision-making in human-AI teams. Recognised with a 2020 Australasian AI Award, she is on the AI100 Standing Committee and has advised national research projects. Currently, she chairs the National Research Infrastructure Advisory Group.
\end{IEEEbiography}
\vspace{-50pt}
\begin{IEEEbiography}[{\includegraphics[width=0.8in,height=1.25in,clip,keepaspectratio]{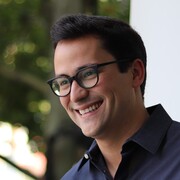}}]{Eduardo Velloso} Eduardo is an interaction technologist and a Professor of Computer Science at the University of Sydney. His research explores the fusion of novel input and sensing modalities to create innovative user experiences. He has a blend of technical expertise with interdisciplinary insights from Engineering, Design, and Psychology. 
\end{IEEEbiography}
\vspace{-60pt}
\begin{IEEEbiography}[{\includegraphics[width=0.8in,height=1.25in,clip,keepaspectratio]{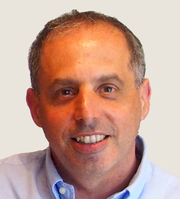}}]{Frank Vetere} Frank Vetere is a Professor of Human-Computer Interaction (HCI) and Deputy Dean (Engagement) at the University of Melbourne. Frank's expertise is in Human-Computer Interaction, with research interests in human-centred approaches to AI, interactivity in mixed reality and technologies for ageing well. 
\end{IEEEbiography}
\vspace{-60pt}
\begin{IEEEbiography}[{\includegraphics[width=0.8in,height=1.25in,clip,keepaspectratio]{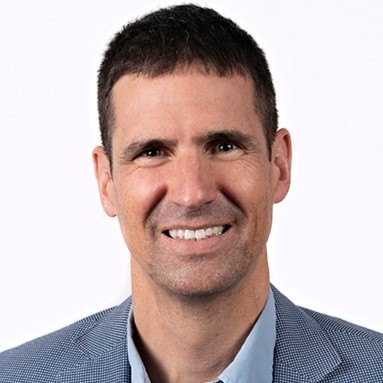}}]{Piers Howe} Piers Howe is an Associate Professor of Psychology at the University of Melbourne. He is known for his expertise in cognitive psychology and human perception. Howe's work spans various aspects of cognitive psychology, including visual perception, attention, and decision-making processes. 
\end{IEEEbiography}
\vspace{-60pt}
\begin{IEEEbiography}[{\includegraphics[width=0.8in,height=1.25in,clip,keepaspectratio]{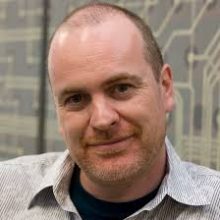}}]{Paul Dourish} Paul Dourish is Chancellor's Professor of Informatics and Associate Dean for Research in the Donald Bren School of Information and Computer Sciences at UC Irvine, and an Honorary Senior Fellow in Computing and Information Systems at the University of Melbourne. His research focuses primarily on understanding information technology as a site of social and cultural production; his work combines topics in human-computer interaction, social informatics, and science and technology studies.
\end{IEEEbiography}

\end{document}